\documentstyle[psfig,11pt,axodraw]{article}
\def\gsim{\:\raisebox{-0.5ex}{$\stackrel{\textstyle>}{\sim}$}\:}
\def\eq#1{{eq. (\ref{#1})}}

\def\np#1#2#3{           {\it Nucl. Phys. }{\bf #1} (19#2) #3}

\def\n.c.#1#2#3{         {\it Nuovo Cim. }{\bf #1} (19#2) #3}
\def\r.n.c.#1#2#3{       {\it Riv. del Nuovo Cim. }{\bf #1} (19#2) #3}

\def\be{\begin{equation}}       
\def\ee{\end{equation}}
\def\bear{\be\begin{array}}
\def\eear{\end{array}\ee}
\def\bea{\begin{eqnarray}}
\def\eea{\end{eqnarray}}

\def\21{$SU(2) \ot U(1)$}
\def\ot{\otimes}
\def\ie{{\it i.e.}}
\def\etal{{\it et al.}}
\def\half{{\textstyle{1 \over 2}}}

\def\quarter{{\textstyle{1 \over 4}}}

\def\eighth{{\textstyle{1 \over 8}}}
\def\bold#1{\setbox0=\hbox{$#1$}
     \kern-.025em\copy0\kern-\wd0
     \kern.05em\copy0\kern-\wd0
     \kern-.025em\raise.0433em\box0 }
\begin{document}
\begin{titlepage}
\begin{flushright}
FTUV/98-54\\
IFIC/98-55\\
hep-ph/9808412\\
\end{flushright}
\vspace*{5mm}
\begin{center} 

{\Large \bf 
Charged Higgs Mass Bounds from {$\bold{b\rightarrow s\gamma}$} 
in a Bilinear   R-Parity Violating Model}\\[15mm]

{\large M. A. D\'\i az, E. Torrente-Lujan, and J. W. F. Valle}\\
\hspace{3cm}\\
{\small Departamento de F\'\i sica Te\'orica, IFIC-CSIC, 
Universitat de Val\`encia}\\ 
{\small Burjassot, Val\`encia 46100, Spain\\
http://neutrinos.uv.es}
\hspace{3cm}\\
\end{center}
\vspace{5mm}
\begin{abstract}

The experimental measurement of $B( b \to s\gamma)$ imposes important
constraints on the charged Higgs boson mass in the MSSM.  If squarks
are in the few TeV range, the charged Higgs boson mass in the MSSM
must satisfy $m_{H^{\pm}} \gsim 440$ GeV. For lighter squarks, then
light charged Higgs bosons can be reconciled with $B(b \to s\gamma)$
only if there is also a light chargino. In the MSSM if we impose
$m_{\chi^{\pm}_1}>90$ GeV then we need $m_{H^{\pm}} \gsim 110$ GeV. We
show that by adding bilinear R--Parity violation (BRpV) in the tau
sector, these bounds are relaxed. The bound on
$m_{H^{\pm}}$ in the MSSM--BRpV model is $\gsim 340$ GeV for the the
heavy squark case and $m_{H^{\pm}} \gsim 75$ GeV for the case of light
squarks. In this case the charged Higgs bosons would be observable at
LEP II. The relaxation of the bounds is due mainly to the fact that
charged Higgs bosons mix with staus and they contribute importantly to
$B(b \to s\gamma)$.

\end{abstract}

\end{titlepage}

\setcounter{page}{1}

{\bf 1.} The first measurement of the inclusive rate for the radiative
penguin decay $b \to s\gamma$ has opened an important window for
physics beyond the Standard Model (SM). The CLEO Collaboration has
reported $B(b \to s\gamma)=(2.32\pm 0.57\pm 0.35)\times 10^{-4}$,
where the first error is statistical and the second is systematic.
Conservatively they find $1.0\times 10^{-4}<B(b \to s\gamma)<4.2\times
10^{-4}$ at $95\%$ C.L. \cite{CLEO}. 
Recently new results have been presented, the new bounds 
are $2.0\times 10^{-4}<B(b \to s\gamma)<4.5\times
10^{-4}$ at $95\%$ C.L. \cite{CLEO2}. 
This measurement has established
for the first time the existence of one--loop penguin diagrams. In
addition, this inclusive branching ratio has been measured by the
ALEPH Collaboration at LEP to be $B(b \to
s\gamma)=(3.11\pm0.80\pm0.72)\times10^{-4}$ \cite{ALEPH}, consistent
with CLEO.

In the SM, loops including the $W$ gauge boson and the unphysical
charged Goldstone boson $G^{\pm}$ contribute to the decay rate. The
latest estimate of this decay rate in the SM is $B(b \to s \gamma) =
(3.28\pm 0.33)\times 10^{-4}$ \cite{CMisiakM}.  This prediction is in
agreement with the CLEO measurement at the $2\sigma$ level.

In two Higgs doublet models (2HDM) the physical charged Higgs boson
$H^{\pm}$ also contributes to the decay rate. In 2HDM of type I, one
Higgs doublet gives mass to the fermions while the other Higgs doublet
decouples from fermions. On the contrary, in 2HDM of type II, the
Higgs doublet $H_1$ gives mass to the down-quarks and the second Higgs
doublet $H_2$ gives mass to the up-quarks. Important constraints on
the charged Higgs mass $m_{H^{\pm}}$ are obtained in 2HDM type II
because the charged Higgs contribution always adds to the SM
contribution \cite{HewettBBP,2HDM}.
Constraints on $m_{H^{\pm}}$ are not important in 2HDM type I because
charged Higgs contributions can have either sign.

In supersymmetric models, loops containing charginos/squarks, 
neutralinos/squarks, and gluino/squarks have to be included \cite{BBMR}.
In the limit of very heavy super-partners, the stringent bounds on 
$m_{H^{\pm}}$ are valid in the Minimal Supersymmetric Standard Model 
(MSSM) because its Higgs sector is of type II \cite{HewettBBP}. 
Nevertheless, even in this case the bound is relaxed at large $\tan\beta$
due to two-loop effects \cite{bsgDiaz1}. It was shown also that by 
decreasing the squarks and chargino masses this bound disappears because 
the chargino contribution can be large and can have the opposite sign to 
the charged Higgs contribution, canceling it \cite{BarbieriG,BSG}. 
Further studies have been made in the MSSM and in its Supergravity version 
\cite{bsgLater,bsgRecent}. As a result, for example, most of the parameter 
space in MSSM-SUGRA is ruled out for $\mu<0$ especially for large 
$\tan\beta$.

QCD corrections are very important and can be a substantial fraction
of the decay rate. Recently, several groups have completed the
Next--to--Leading order QCD corrections to $B(b \to s \gamma)$.
Two--loop corrections to matrix elements were calculated in
\cite{GHW}. The two--loop boundary conditions were obtained in
\cite{AY} (see also \cite{CDGG}). Bremsstrahlung corrections were 
obtained in \cite{bremss}. Finally, three--loop anomalous dimensions
in the effective theory used for resumation of large logarithms
$\ln(m_W^2/m_b^2)$ were found in \cite{CMisiakM,anomalous} (see also
\cite{anomOther}). In this work we include all these QCD corrections.

All previous work on $b \to s\gamma$ in supersymmetry has assumed the
conservation of R--Parity. Here, we introduce in the superpotential of
the MSSM the term $\epsilon_3\widehat L_3\widehat H_2$, which violates
R--Parity and tau--lepton number explicitly. This is motivated by
models where R--Parity is broken spontaneously \cite{SRpSB,SRpSB2}
through a right handed sneutrino vacuum expectation value. This in
turn induces Bilinear R--Parity Violation (BRpV)
\cite{e3others,epsrad,v3cha,othersUs}.

Relative to the calculation of $B(b \to s\gamma)$, the main difference
of these models with respect to the MSSM is that in BRpV the charged
Higgs boson mixes with the staus and the tau lepton mixes with the
charginos.  This way, new contributions have to be added and the old
contributions are modified by mixing angles. In this paper we study
how the BRpV model affects the $B(b \to s\gamma)$ prediction and, in
particular, the bounds on the charged Higgs mass derived from the
experimental constraint on the branching ratio of this decay. In the
numerical part of our work, we do not embed our model into SUGRA
scenarios, rather we consider all unknown parameters to be free at the
weak scale, and we call this model unconstrained MSSM--BRpV.


{\bf 2.} The superpotential we consider here contains the following 
bilinear terms
\begin{equation} 
W_{Bi}=\varepsilon_{ab}\left[
-\mu\widehat H_1^a\widehat H_2^b
+\epsilon_3\widehat L_3^a\widehat H_2^b\right]\,,
\label{eq:WBi}
\end{equation}
where both parameters $\mu$ and $\epsilon_3$ have units of mass, and the
last one violates R--Parity and tau--lepton number.

One of the main characteristics of BRpV is that a tau sneutrino vacuum
expectation value (vev) $v_3$ is induced. The $v_3$ is related to the
mass parameter $\epsilon_3$ through a minimization condition. This
non--zero sneutrino vev is present even in a basis where the
$\epsilon_3$ term disappears from the superpotential.  This basis is
defined by the rotation $\mu'\widehat H_1'=\mu\widehat
H_1-\epsilon_3\widehat L_3$ and $\mu'\widehat L_3'=\epsilon_3\widehat
H_1+\mu\widehat L_3$, where we have set $\mu'^2=\mu^2+\epsilon_3^2$.
The sneutrino vev in this basis, which we denote by $v'_3$, is
non--zero due to mixing terms that appear in the soft sector between
$\widetilde L_3$ and $H_1$ scalars.
It is also possible to choose a basis where the sneutrino vev is
zero. In this basis a non--zero $\epsilon_3$ term is present in the
superpotential \cite{BKMO}. All three basis are equivalent.

In addition, a mixing between neutralinos and the tau neutrino is induced.
In the {\sl rotated} basis, the neutralino/neutrino mass matrix reads
\begin{equation} 
{\bold M}_N=\left[  
\begin{array}{ccccc}  
M^{\prime } & 0 & -\frac 12g^{\prime }v'_1 & \frac 12g^{\prime }v_2 & 
-\frac  12g^{\prime }v'_3 \\   
0 & M & \frac 12gv'_1 & -\frac 12gv_2 & \frac 12gv'_3 \\   
-\frac 12g^{\prime }v'_1 & \frac 12gv'_1 & 0 & -\mu'  & 0 \\   
\frac 12g^{\prime }v_2 & -\frac 12gv_2 & -\mu'  & 0 & 0 \\   
-\frac 12g^{\prime }v'_3 & \frac 12gv'_3 & 0 & 0 & 0  
\end{array}  
\right] 
\label{eq:NeuM5x5rot} 
\end{equation} 
where $M$ and $M'$ are the gaugino soft breaking masses. In
\eq{eq:NeuM5x5rot} the last column and row correspond to the rotated
sneutrino field. Clearly, a tau neutrino mass is induced and is
proportional to $v'^2_3$.  The experimental bound on the tau neutrino
mass, given by $m_{\nu_{\tau}}<18$ MeV \cite{tauBound}, implies an
upper bound for $v'_3$ of about 5--10 GeV. Cosmological bounds are
stronger and have been discussed in ref.~\cite{desert}. Considering
that $v'_3=(\epsilon_3 v_1+\mu v_3)/\mu'$, this may seem a fine
tuning, nevertheless, it is not so. The reason is that in models with
universality of soft mass parameters at the unification scale, $v'_3$
is radiatively generated and is proportional to the bottom quark
Yukawa coupling squared.  In this way, $v'_3$ as well $m_{\nu_{\tau}}$
are {\sl calculable} and naturally small \cite{epsrad}.

{\bf 3.} In BRpV, the tau lepton mixes with the charginos forming a
set of three charged fermions $F_i^{\pm}$, $i=1,2,3$. In the
{\sl{original}} basis where $\psi^{+T}=(-i\lambda^+,\widetilde
H_2^1,\tau_R^+)$ and $\psi^{-T}=(-i\lambda^-,\widetilde
H_1^2,\tau_L^-)$, the charged fermion mass terms in the Lagrangian are
${\cal L}_m=-\psi^{-T}{\bold M_C}\psi^+$, with the mass matrix given
by
\begin{equation} 
{\bold M_C}=\left[\matrix{ 
M & {\textstyle{1\over{\sqrt{2}}}}gv_2 & 0 \cr 
{\textstyle{1\over{\sqrt{2}}}}gv_1 & \mu &  
-{\textstyle{1\over{\sqrt{2}}}}h_{\tau}v_3 \cr 
{\textstyle{1\over{\sqrt{2}}}}gv_3 & -\epsilon_3 & 
{\textstyle{1\over{\sqrt{2}}}}h_{\tau}v_1}\right] 
\label{eq:ChaM3x3} 
\end{equation} 
and where $h_{\tau}$ is the tau Yukawa coupling. Note that in BRpV the 
relation between $h_{\tau}$ and $m_{\tau}$ is different than in the MSSM
due to the mixing with charginos \cite{v3cha}. In this way, in BRpV not
only the charginos contribute to $b \to  s\gamma$ but also the tau 
lepton. 

In the notation of ref.~\cite{BarbieriG} we have that the chargino/tau 
amplitude is
\begin{eqnarray}
A_{\gamma,g}^{F^{\pm}}&=&\sum_{i=1}^3\Bigg\{
{{m_W^2}\over{m_{F_i^{\pm}}^2}}\bigg[|V_{i1}|^2f^{(1)}
\Big({{m_{\tilde q}^2}\over{m_{F_i^{\pm}}^2}}\Big)-\sum_{j=1}^2
\bigg|V_{i1}R_{\tilde t}^{j1}-V_{i2}R_{\tilde t}^{j2}
{{m_t}\over{\sqrt{2}m_Ws_{\beta}s_{\theta}}}\bigg|^2
f^{(1)}\Big({{m_{\tilde t_j}^2}\over{m_{F_i^{\pm}}^2}}\Big)
\nonumber\\\label{Charginobsg}\\
&&\!\!\!\!\!\!\!\!\!\!\!\!\!\!\!\!\!\!\!\!
-{{U_{i2}}\over{\sqrt{2}c_{\beta}s_{\theta}}}
{{m_W}\over{m_{F_i^{\pm}}}}\bigg[V_{i1}f^{(3)}
\Big({{m_{\tilde q}^2}\over{m_{F_i^{\pm}}^2}}\Big)-\sum_{j=1}^2
\bigg(V_{i1}R_{\tilde t}^{j1}-V_{i2}R_{\tilde t}^{j2}
{{m_t}\over{\sqrt{2}m_Ws_{\beta}s_{\theta}}}\bigg)R_{\tilde t}^{j1}
f^{(3)}\Big({{m_{\tilde t_j}^2}\over{m_{F_i^{\pm}}^2}}\Big)
\bigg]\Bigg\}
\nonumber
\end{eqnarray}
where the sum goes from one to three, in order to account for the chargino
and the tau lepton contributions. The matrices $V$ and $U$ are $3\times 3$
and diagonalize the chargino/tau mass matrix in \eq{eq:ChaM3x3} 
according to
\begin{equation}
{\bf U}^*{\bf M_C}{\bf V}^{-1}=\left[\matrix{
m_{\chi^{\pm}_1} & 0 & 0 \cr
0 & m_{\chi^{\pm}_2} & 0 \cr
0 & 0 & m_{\tau}}\right]\,.
\label{eq:ChaMdiag}
\end{equation}
with $m_{\chi^{\pm}_1}<m_{\chi^{\pm}_2}$. The value of $h_{\tau}$ is
fixed by the condition $m_{\tau}=1.777$ GeV as a function of SUSY
parameters. The matrix $R_{\tilde t}$ is the rotation matrix which
diagonalizes the stop quark mass matrix \cite{epsrad} necessary to
take into account the left--right mixing in the stop mass matrix. We
neglect this mixing for the other up--type squarks.  Finally, in
\eq{Charginobsg} we have defined the angles $\beta$ and $\theta$
in spherical coordinates
\begin{equation}
v_1=v\cos\beta\sin\theta\,,\qquad v_2=v\sin\beta\sin\theta\,,\qquad
v_3=v\cos\theta\,,
\label{vevdef}
\end{equation}
where $v=246$ GeV and the MSSM relation $\tan\beta=v_2/v_1$ is preserved.

In order to study the effect of BRpV on $B(b \to s\gamma)$ we make a
scan over parameter space which contains over $5\times 10^4$
points. We have varied randomly the parameters in the following ranges:
\begin{eqnarray}
&\mid \mu,B\mid          &<500 \,\,{\mathrm{GeV}}\,,
\nonumber\\
 0.5<&\tan\beta      &<30 \,,
\nonumber\\
  10<&M_{L_3},M_{R_3}&<1000 \,\,{\mathrm{GeV}}\,,
\nonumber\\
 100<&M_Q=M_U        &<1500 \,\,{\mathrm{GeV}}\,,
\nonumber\\
  50<&M = 2M'          &<1000 \,\,{\mathrm{GeV}}\,,
\nonumber\\
&\mid A_t,A_{\tau}\mid   &<500 \,\,{\mathrm{GeV}}
\label{paramMSSM}
\end{eqnarray}
for the MSSM parameters, and
\begin{eqnarray}
&\mid\epsilon_3\mid &<200 \,\,{\mathrm{GeV}}\,,
\nonumber\\
&\mid v'_3   \mid    &<10 \,\,{\mathrm{GeV}}
\label{paramBRpV}
\end{eqnarray}
for the BRpV parameters. In \eq{paramMSSM}, $B$ is the bilinear soft
mass parameter associated with the $\mu$ term in the superpotential,
$M_{L_3}$ and $M_{R_3}$ are the soft mass parameters in the stau
sector, $M_Q$ and $M_U$ are the soft mass parameters in the stop
sector. The parameters $A_t$ and $A_{\tau}$ are the trilinear soft
masses in the stop and stau sector respectively. Note that $B_2$, the
bilinear soft mass parameter associated with the $\epsilon_3$ term in
the superpotential, is fixed by the minimization equations of the
scalar potential.

\begin{figure}
\centerline{\protect\hbox{\psfig{file=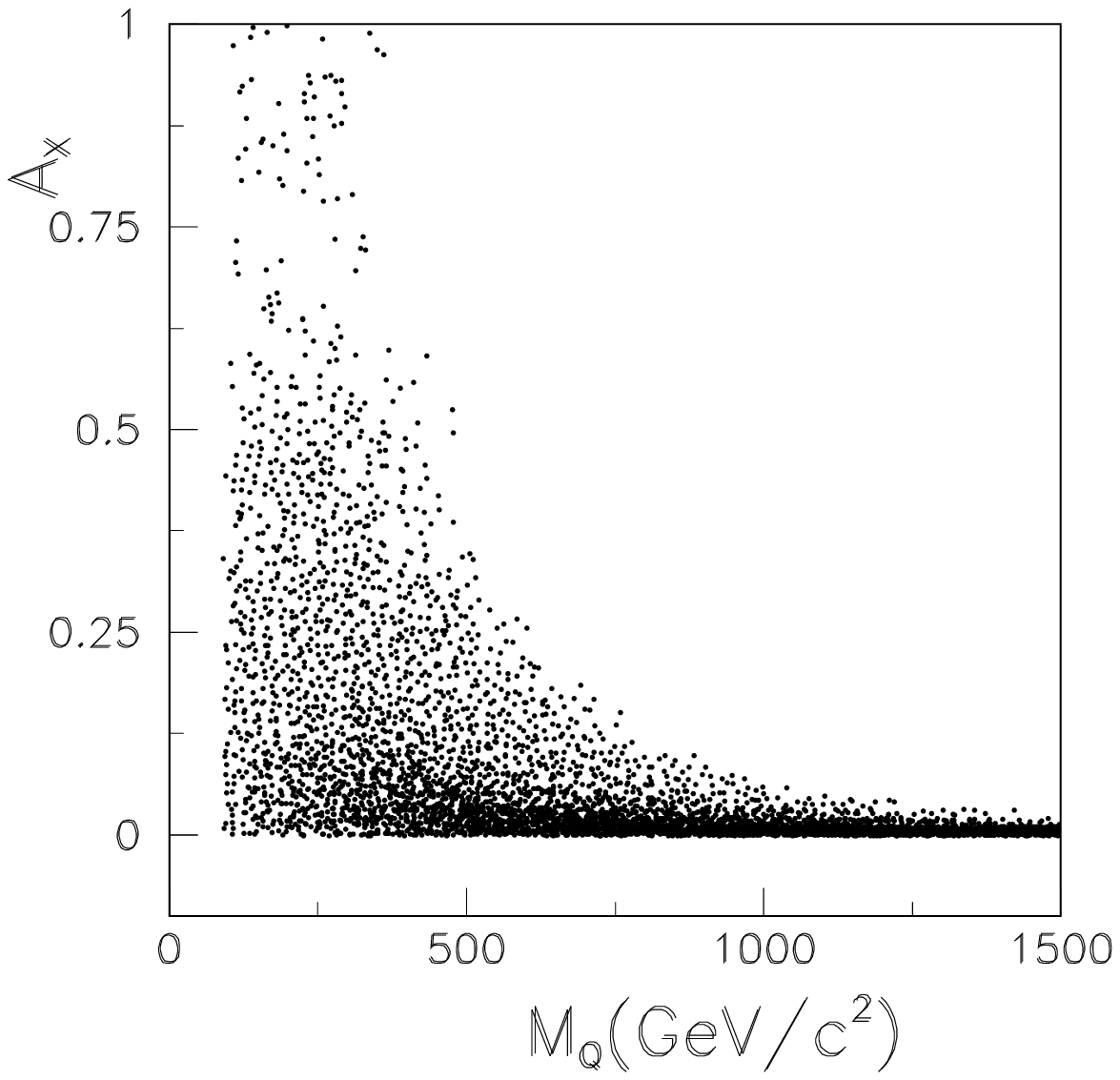,height=9cm,width=0.6\textwidth}}}
\caption{Chargino/tau amplitude contributing to $B(b \to  s\gamma)$ 
as a function of the squark soft mass parameter $M_Q$ in MSSM--BRpV.}
\label{chamq}
\end{figure}

The amplitude $A_{\gamma}^{F^{\pm}}$ is plotted in Fig.~\ref{chamq} as
a function of the soft breaking squark mass parameter $M_Q$. One can
clearly see that the $A_{\gamma}^{F^{\pm}}$ contribution falls as the
squark mass increases. For $M_Q=1.5$ TeV the maximum chargino
amplitude goes down to 2\%.

\begin{figure}
\centerline{\protect\hbox{\psfig{file=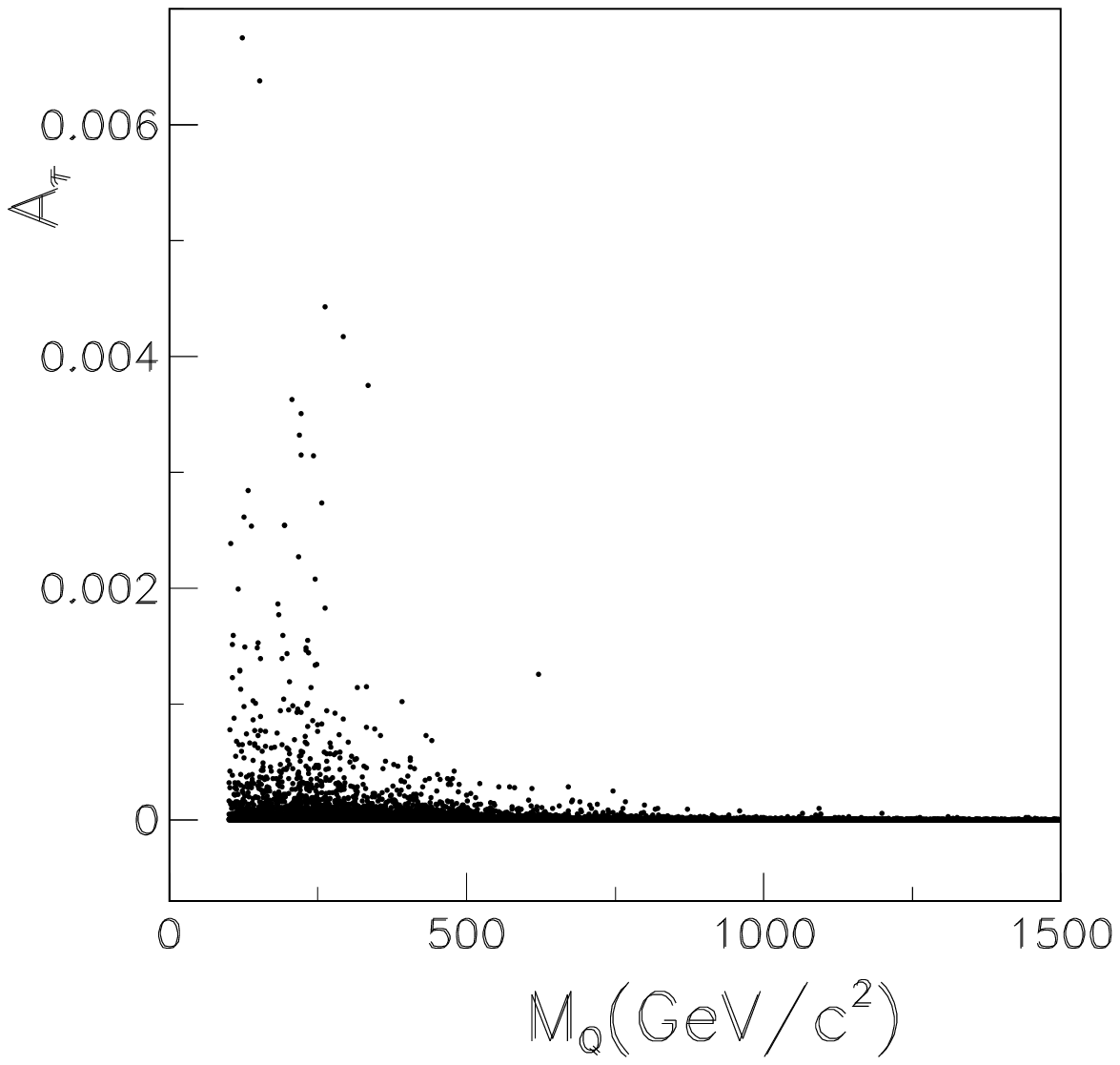,height=9cm,width=0.6\textwidth}}}
\caption{Tau amplitude contributing to $B(b \to  s\gamma)$ 
as a function of the squark soft mass parameter $M_Q$ in MSSM--BRpV.}
\label{taumq}
\end{figure}

In order to appreciate the relative importance of the tau contribution
to $B(b \to s\gamma)$ in \eq{Charginobsg}, we have plotted this
amplitude in Fig.~\ref{taumq}. Clearly, the tau contribution can be
neglected since it is less than 0.6\% of the total. This can be
understood as follows. First, the tau lepton contributions to the
second line in \eq{Charginobsg} are small due to the small tau mass,
since the function $f^{(3)}(x)$ satisfies $\sqrt{x}f^{(3)}(x) \to 0$
as $x \to \infty$.  On the other hand the tau contributions to the
first line are small because the right-handed tau does not mix
appreciably with the Higgsino, implying that $V_{31}$ and $V_{32}$ are
small. 

Let us also note that we in the above figures we have implemented the
LEP bound on the lightest chargino mass of 90 GeV. Strictly speaking,
the chargino mass bound in the MSSM does not directly apply to BRpV
but we do not expect any sizeable relaxation of the bound. For a
recent analysis see ref. \cite{tauj}.

{\bf 4.} We now turn to our main results. In the MSSM--BRpV, the
charged Higgs sector mixes with the stau sector forming a set of four
charged scalars. The mass terms in the scalar potential are given by
$V_{quadratic}={\bf\Phi}^-{\bf M}_{S^{\pm}}^2{\bf\Phi}^{+T}$, where
${\bf\Phi}^{\pm}=(H_1^{\pm},H_2^{\pm},\tilde\tau_L^{\pm},
\tilde\tau_R^{\pm})$ are the fields in the original basis.

The $4\times4$ charged scalar mass matrix ${\bf M}_{S^{\pm}}^2$ has
been studied in detail in ref.~\cite{v3cha}. It is diagonalized by a
rotation matrix ${\bf R}_{S^{\pm}}$ such that the eigenvectors are
${\bf S}^{\pm}={\bf R}_{S^{\pm}}{\bf\Phi}^{\pm}$, and the eigenvalues
are
${\mathrm{diag}}(m_{G^{\pm}}^2,m_{H^{\pm}}^2,m_{\tilde\tau_1}^2,m_{\tilde
\tau_2}^2)={\bf R}_{S^{\pm}}{\bf M}_{S^{\pm}}^2{\bf R}_{S^{\pm}}^{\dagger}$. 
Of course, the massless eigenvector is associated to the unphysical charged 
Goldstone boson $G^{\pm}$ eaten by the $W$ boson.

In contrast, if we work in the basis where the $\epsilon_3$ term
disappears from the superpotential (described earlier), then the mass
terms of the charged scalar sector are given by
$V_{quadratic}={\bf\Phi'}^-{\bf M'}_{S^{\pm}}^2{\bf\Phi'}^{+T}$ with
${\bf\Phi'}^{\pm}=(H'^{\pm}_1,H_2^{\pm},\tilde\tau'^{\pm}_L,
\tilde\tau_R^{\pm})$, and the charged scalar mass matrix reads
\begin{equation} 
\bold{M'^2_{S^{\pm}}}=\left[\matrix{ 
{\bold M'^2_H} & {\bold M'^{2T}_{H\tilde\tau}} \cr 
{\bold M'^2_{H\tilde\tau}} & {\bold M'^2_{\tilde\tau}} 
}\right]\,,
\label{eq:subdivM} 
\end{equation} 
In \eq{eq:subdivM} we have decomposed the mass matrix in $2\times2$
blocks. The charged Higgs sector is
\begin{equation} 
{\bold M'^2_H}=\left[\matrix{ 
m'^2_{H_1}+\mu'^2+D+\half h_{\tau}^2v'^2_3+\quarter g^2(v_2^2-v'^2_3)
& B'\mu'+\quarter g^2v'_1v_2 
\cr B'\mu'+\quarter g^2v'_1v_2 
& m^2_{H_2}+\mu'^2-D+\quarter g^2(v'^2_1+v'^2_3)
}\right] 
\label{eq:subMHH}
\end{equation} 
with $D=\eighth(g^2+g'^2)(v'^2_1-v_2^2+v'^2_3)$ and the vacuum
expectation values of the fields $H'_1$ and $L'_3$ satisfying
$v'_1=(\mu v_1-\epsilon_3 v_3)/\mu'$ and $v'_3=(\epsilon_3 v_1+\mu
v_3)/\mu'$ respectively. The soft mass parameters which appear in the
new basis are related to the original soft mass parameters according
to
\begin{equation}
m'^2_{H_1}={{m_{H_1}^2\mu^2+M_{L_3}^2\epsilon_3^2}\over{\mu'^2}}\,,\quad
M'^2_{L_3}={{m_{H_1}^2\epsilon_3^2+M_{L_3}^2\mu^2}\over{\mu'^2}}\,,\quad
B'={{B\mu^2+B_2\epsilon_3^2}\over{\mu'^2}}\,.
\label{newmasses}
\end{equation}
The $2\times2$ stau sub-matrix is given by
\begin{equation} 
{\bold M'^2_{\tilde\tau}}=\left[\matrix{ 
M'^2_{L_3}+\half h_{\tau}^2v'^2_1+D-\quarter g^2(v'^2_1-v_2^2)
& {1\over{\sqrt{2}}}h_{\tau}(A_{\tau}v'_1-\mu' v_2) 
\cr {1\over{\sqrt{2}}}h_{\tau}(A_{\tau}v'_1-\mu' v_2) 
& M_{R_3}^2+\half h_{\tau}^2(v'^2_1+v'^2_3)-D' 
}\right] 
\label{eq:subtautau}
\end{equation} 
where $D'=\quarter g'^2(v'^2_1-v_2^2+v'^2_3)$. Finally, the Higgs--stau
mixing is
\begin{equation} 
{\bold M'^2_{H\tilde\tau}}=\left[\matrix{ 
\mu\epsilon_3\Delta m^2/\mu'^2-\half h_{\tau}^2v'_1v'_3
+\quarter g^2v'_1v'_3 
& -\mu\epsilon_3\Delta B/\mu'+\quarter g^2v_2v'_3 
\cr -{1\over{\sqrt{2}}}h_{\tau}A_{\tau}v'_3 
& -{1\over{\sqrt{2}}}h_{\tau}\mu' v'_3
}\right] 
\label{eq:subHtau} 
\end{equation} 
where $\Delta m^2=m_{H_1}^2-M_{L_3}^2$ and $\Delta B=B_2-B$ indicate how
much deviation from universality of soft masses we have at the weak scale.
The mass matrix is diagonalized by a rotation matrix ${\bf R'}_{S^{\pm}}$.

In models where MSSM--BRpV is embedded into Supergravity \cite{epsrad}
and universality of soft masses is assumed at the unification scale, 
$\Delta m^2$ and $\Delta B$ are calculable, one--loop induced, and 
proportional to the square of the bottom quark Yukawa coupling. In
addition, imposing that the vacuum expectation values are solutions
of the minimization of the scalar potential we find that the following
tadpole equations associated to the rotated Higgs fields must hold:
\begin{eqnarray}
&&\mu'^2v'_1+m'^2_{H_1}v'_1-B'\mu'v_2+\Delta m^2
{{\epsilon_3\mu}\over{\mu'^2}}v'_3+Dv'_1=0
\nonumber\\
&&\mu'^2v_2+m^2_{H_2}v_2-B'\mu'v'_1+\Delta B
{{\epsilon_3\mu}\over{\mu'}}v'_3+Dv_2=0
\label{tadpoleiyii}
\end{eqnarray}
together with the equation associated to the rotated sneutrino field:
\begin{equation}
\Delta m^2{{\epsilon_3\mu}\over{\mu'^2}}v'_1
+\Delta B{{\epsilon_3\mu}\over{\mu'}}v_2+M'^2_{L_3}v'_3+Dv'_3=0\,.
\label{tadpoleiii}
\end{equation}
{}From this last equation we see that in SUGRA--BRpV the vacuum
expectation value $v'_3$ is also small and proportional to the bottom
quark Yukawa coupling squared, proving that the tau neutrino mass is
naturally small.  As mentioned before, in our scan we work with the
unconstrained MSSM--BRpV, where all the parameters are free at the
weak scale. Of course, in order to have a correct electroweak symmetry
breaking, we impose the tadpole equations in eqs.~(\ref{tadpoleiyii})
and (\ref{tadpoleiii}). In addition, we enforce the experimental tau
neutrino mass upper limit $m_{\tau}<18$ MeV (our results are not
changed if we impose $m_{\tau}<1$ MeV instead). Barring cancellations
in \eq{tadpoleiii} the $\nu_tau$ constraint restricts the terms
proportional to $\Delta m^2$ and $\Delta B$ to be small.

It is clear from \eq{eq:subHtau} that the Higgs--stau mixing,
defined in the rotated basis, vanishes in the limit $v'_3 \to  0$.
Therefore, charged Higgs and stau sectors defined in this basis decouple 
from each other. In addition, in this limit, \eq{eq:subMHH} and 
\eq{eq:subtautau} reduce to MSSM--looking charged Higgs and stau 
mass matrices. Motivated by this, we can define the charged Higgs as the 
massive field $S_i^{\pm}$ with largest component along $H'^{\pm}_1$ and 
$H'^{\pm}_2$, \ie, maximum 
$({\bf R'}_{S^{\pm}}^{i1})^2+({\bf R'}_{S^{\pm}}^{i2})^2$.

On the other hand, consider the charged scalar couplings to top and 
bottom quarks, which are equal to
\begin{center}
\vspace{-60pt} \hfill \\
\begin{picture}(120,90)(0,21) 
\DashLine(10,25)(60,25){5}
\Vertex(60,25){2}
\ArrowLine(110,-5)(60,25)
\ArrowLine(60,25)(110,55)
\Text(85,0)[]{$b$}
\Text(85,51)[]{$t$}
\Text(35,38)[]{$S^+_i$}
\end{picture}
$
=i\lambda_i^LP_L+i\lambda_i^RP_R
$
\vspace{30pt} \hfill \\
\end{center}
%
where $\lambda_i^L=R_{S^{\pm}}^{ij}\lambda^{0L}_j$ with 
$\lambda^{0L}=(0,h_t,0,0)$, and $\lambda_i^R=R_{S^{\pm}}^{ij}\lambda^{0R}_j$ 
with $\lambda^{0R}=(h_b,0,0,0)$. These couplings reflex the fact that
only $H_1^{\pm}$ and $H_2^{\pm}$, and not weak staus, couple to quarks. 
Therefore, another motivated definition is to call the charged Higgs as 
the massive field $S_i^{\pm}$ with largest component along $H^{\pm}_1$ 
and $H^{\pm}_2$, \ie, maximum 
$({\bf R}_{S^{\pm}}^{i1})^2+({\bf R}_{S^{\pm}}^{i2})^2$.
Both definitions coincide in the limit $\epsilon_3 \to  0$ and
are equally good. This ambiguity present in the case of non--negligible
$\epsilon_3$ is simply due to the fact that now we have a set of four
charged scalars which are a mixture of Higgs and staus. In this paper we 
have worked with both definitions.


{\bf 5.} The charged scalar amplitude contributing to 
$B(b \to  s\gamma)$ is
\begin{equation}
A_{\gamma,g}^{S^{\pm}}={1\over 2}\sum_{i=2}^4
{{m_t^2}\over{m_{S^{\pm}_i}^2}}\bigg[
{1\over{s_{\beta}^2s_{\theta}^2}}\Big({\bf R}_{S^{\pm}}^{i2}\Big)^2
f^{(1)}\Big({{m_t^2}\over{m_{S^{\pm}_i}^2}}\Big)+
{1\over{s_{\beta}c_{\beta}s_{\theta}^2}}
\Big({\bf R}_{S^{\pm}}^{i1}{\bf R}_{S^{\pm}}^{i2}\Big)
f^{(2)}\Big({{m_t^2}\over{m_{S^{\pm}_i}^2}}\Big)\bigg]\,.
\label{ChSbsg}
\end{equation}
The first charged scalar ($i=1$) corresponds to the unphysical charged 
Goldstone boson which contributes to the SM amplitude, thus it is omitted. 
In BRpV, three scalars are contributing to the 
amplitude in \eq{ChSbsg}: the charged Higgs boson and the two 
staus. The charged scalar couplings to quarks, which multiply each 
function $f^{(i)}$ involve the matrix ${\bf R}_{S^{\pm}}$ which 
diagonalizes the charged scalar mass matrix in the {\sl unrotated} 
basis. 

\begin{figure}
\centerline{\protect\hbox{\psfig{file=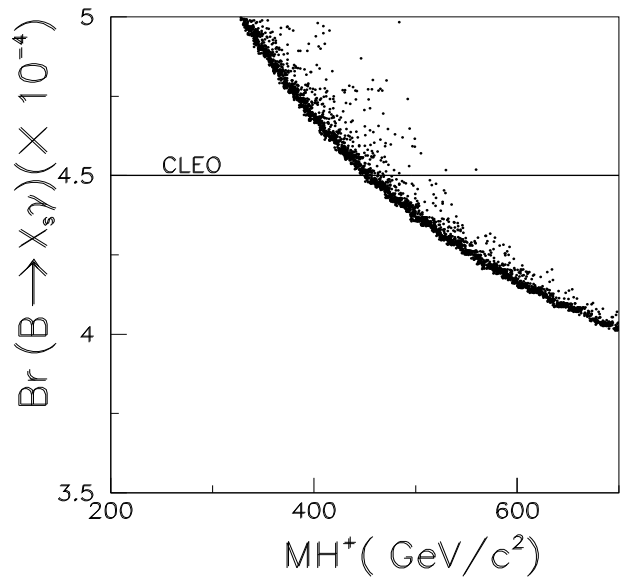,height=9cm,width=0.6\textwidth}}}
\caption{Branching ratio $B(b \to  s\gamma)$ as a function of the 
charged Higgs boson mass $m_{H^{\pm}}$ in the limit of very heavy
squark masses within the MSSM.}
\label{brmh1}
\end{figure}

In order to have an idea of the effects of BRpV on the constraints
from the measurement of $B(b \to s\gamma)$ it is instructive to take
the limit of very massive squarks. In this limit the chargino
amplitude in \eq{Charginobsg} can be neglected relative to the charged
scalar amplitude. It is well known that in this scenario a lower limit
on the MSSM charged Higgs mass is inferred. In Fig.~\ref{brmh1} we
plot the branching ratio $B(b \to s\gamma)$ as a function of the
charged Higgs mass $m_{H^{\pm}}$ in the MSSM with large squark masses
(in practice, masses at least equal to several TeV are necessary to
suppress the chargino amplitude). The horizontal dashed line
corresponds to the latest CLEO upper limit and, therefore, a lower limit of
approximately $m_{H^{\pm}}>440$ GeV is found. We note that in
implementing the QCD corrections we simply take the $B$ scale $Q_b=5$
GeV (see ref.~\cite{MPR} for a discussion on the uncertainties of the
QCD corrections to the branching ratio).

\begin{figure}
\centerline{\protect\hbox{\psfig{file=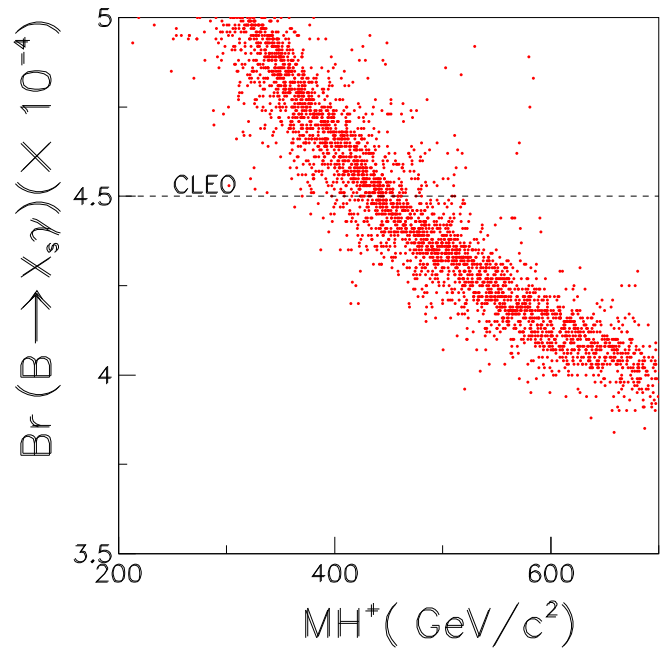,height=9cm,width=0.6\textwidth}}}
\caption{Branching ratio $B(b \to  s\gamma)$ as a function of the 
charged Higgs boson mass $m_{H^{\pm}}$ in the limit of very heavy
squark masses in MSSM--BRpV. The charged Higgs boson is defined as the 
massive charged scalar field with largest couplings to quarks.}
\label{brmh2}
\end{figure}

In Fig.~\ref{brmh2} we plot $B(b \to s\gamma)$ as a function of
$m_{H^{\pm}}$ in the MSSM--BRpV model in the heavy squark limit. The
difference is exclusively due to the mixing of the charged Higgs boson
with the staus.  The bound on the charged Higgs mass is in this case
approximately $m_{H^{\pm}}>340$ GeV. Therefore, the bound is relaxed
by about 100 GeV.  The reason for the relaxation of the bound is
simple. While the charged Higgs couplings to quarks diminish due to
Higgs--Stau mixing, the contribution from the staus does not always
compensate it, because staus may be heavier than the charged Higgs
boson. It is important to stress that in Fig.~\ref{brmh2} we have
defined the charged Higgs as the field $S^{\pm}_i$ (excluding the
massless Goldstone boson) that couples stronger to quarks, \ie, the
massive field which maximizes the quantity 
$({\bf R}_{S^{\pm}}^{i1})^2+({\bf R}_{S^{\pm}}^{i2})^2$. Since in the charged
Higgs loops contributing to $b \to s\gamma$ the relevant couplings are
precisely those, this definition seems to be the most relevant for our
purpose. Nevertheless, in order to compare, we have adopted a second
way to decide which of the charged scalars we define as the charged
Higgs.

\begin{figure}
\centerline{\protect\hbox{\psfig{file=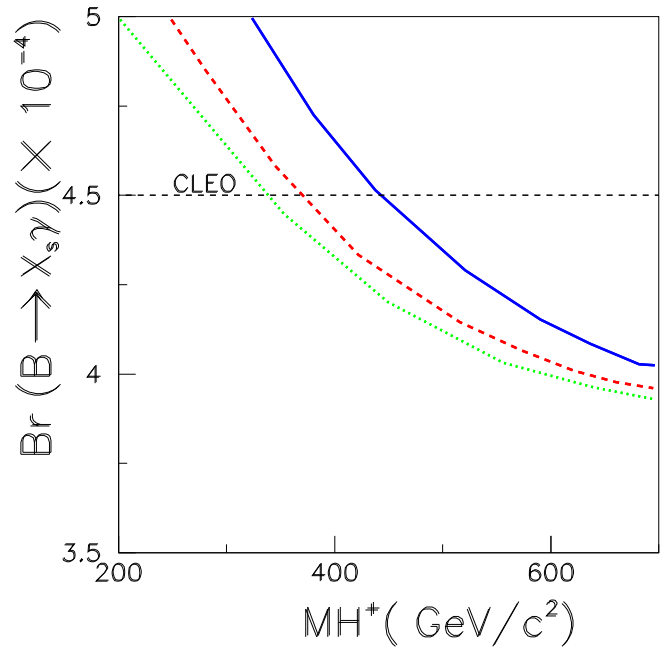,height=9cm,width=0.6\textwidth}}}
\caption{Lower limit on the branching ratio $B(b \to  s\gamma)$ as 
a function of the charged Higgs boson mass $m_{H^{\pm}}$. We consider 
the limit of very heavy squark masses within the MSSM (solid) and 
the MSSM--BRpV (dashes and dots as explained in the text).}
\label{brmh}
\end{figure}

In Fig.~\ref{brmh} we plot lower limits of $B(b \to s\gamma)$ as a
function of $m_{H^{\pm}}$ in the MSSM--BRpV in the heavy squark limit.
In the solid line we have the MSSM limit inferred from
Fig.~\ref{brmh1} and the dotted line is the MSSM--BRpV limit deduced
from Fig.~\ref{brmh2}, where the charged Higgs is defined as the
massive scalar with largest couplings to quarks. An alternative
definition is to consider the charged Higgs boson as the massive field
$S_i^{\pm}$ with largest component along $H'^{\pm}_1$ and
$H'^{\pm}_2$, \ie, maximum $({\bf R'}_{S^{\pm}}^{i1})^2+({\bf
R'}_{S^{\pm}}^{i2})^2$, as already explained in the text. This
definition is motivated by the fact that in the rotated basis, where
the epsilon term disappears from the superpotential, the rotated
charged Higgs fields $H'^{\pm}_1$ and $H'^{\pm}_2$ decouple from the
rotated staus fields as $v'_3 \to 0$.  The corresponding lower limit
of $B(b \to s\gamma)$ is represented by the dashed line in
Fig.~\ref{brmh} and lies between the other two limits.  We observe
that the effect of the relaxation of the bound on $m_{H^{\pm}}$ is
maintained although slightly weaker. The bound from the dashed line in
Fig.~\ref{brmh} is approximately $m_{H^{\pm}}>370$ GeV, implying that
BRpV relaxes the bound by about 70 GeV with respect to the MSSM.

\begin{figure}
\centerline{\protect\hbox{\psfig{file=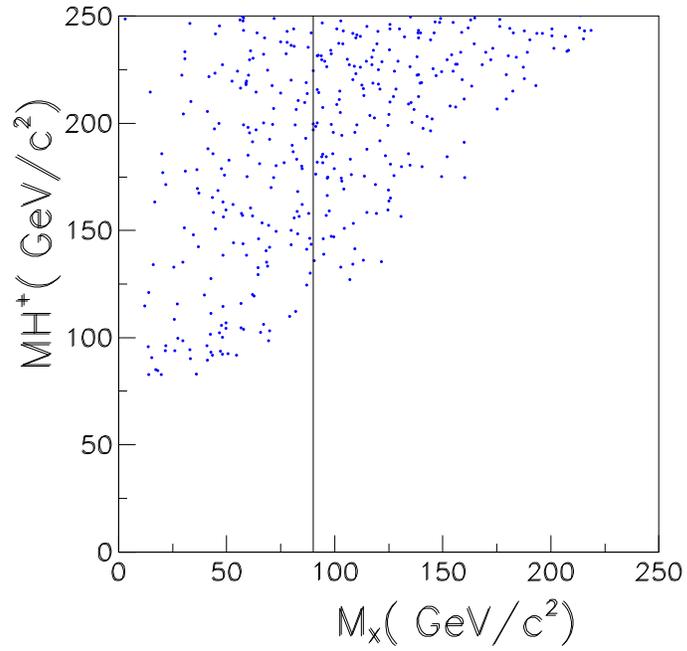,height=9cm,width=0.6\textwidth}}}
\caption{Charged Higgs boson mass as a function of the lightest chargino 
mass for $B(b \to  s\gamma)$ compatible with CLEO measurement within 
the MSSM. The vertical dashed line corresponds to $m_{\chi_1}=90$ GeV.}
\label{mhcha1}
\end{figure}

Another interesting region of parameter space to explore is the region of
light charged Higgs boson and light chargino. It is known that in order 
to have a light charged Higgs boson, its large contribution to 
$B(b \to  s\gamma)$ must be canceled by the contribution from
light charginos and stops. In Fig.~\ref{mhcha1} we plot the charged Higgs 
mass $m_{H^{\pm}}$ as a function of the lightest chargino mass
$m_{\chi^{\pm}_1}$ within the MSSM. All the points satisfy the CLEO
bound mentioned before. The solid vertical line is defined by 
$m_{\chi^{\pm}_1}=90$ GeV, which is approximately the experimental
lower limit found by LEP2, at least for the heavy sneutrino case.
Therefore, we can say that in order to have $m_{\chi^{\pm}_1}>90$ GeV, 
the CLEO measurement of $B(b \to  s\gamma)$ implies that
$m_{H^{\pm}}\gsim 110$ GeV in the MSSM. 

\begin{figure}
\centerline{\protect\hbox{\psfig{file=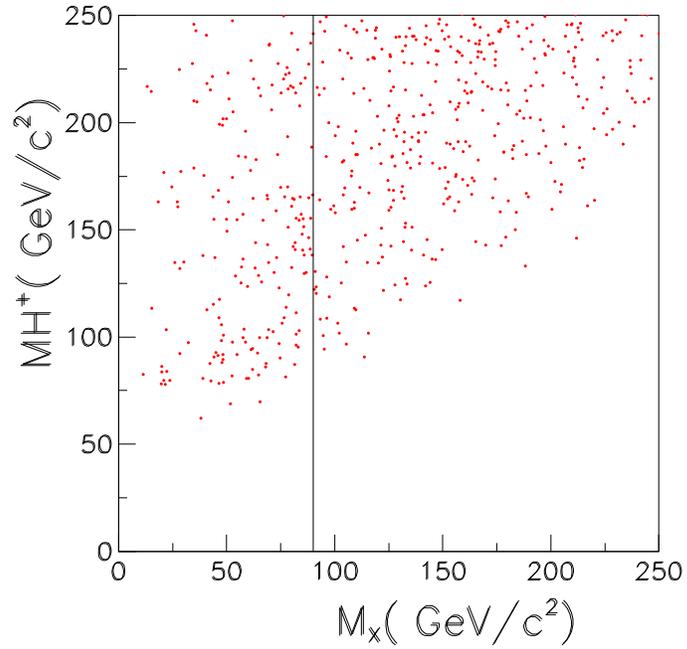,height=9cm,width=0.6\textwidth}}}
\caption{Charged Higgs boson mass as a function of the lightest chargino 
mass for $B(b \to  s\gamma)$ compatible with CLEO measurement in 
MSSM--BRpV. The charged Higgs is defined as the massive charged scalar
field with largest couplings to quarks. The vertical dashed line 
corresponds to $m_{\chi_1}=90$ GeV.}
\label{mhcha2}
\end{figure}

As before, this bound on the charged Higgs boson mass is relaxed in
the MSSM--BRpV model. In Fig.~\ref{mhcha2} we plot $m_{H^{\pm}}$
versus $m_{\chi^{\pm}_1}$ for points satisfying the CLEO bound on $B(b
\to s\gamma)$ within the MSSM--BRpV. The charged Higgs boson is
defined as the massive charged scalar with strongest couplings to
quarks.  We see from Fig.~\ref{mhcha2} that in order to have
$m_{\chi^{\pm}_1}>90$ GeV compatible with $B(b \to s\gamma)$ we need
$m_{H^{\pm}}\gsim 75$ GeV, therefore, relaxing the bound by about 35
GeV with respect to the MSSM.

\begin{figure}
\centerline{\protect\hbox{\psfig{file=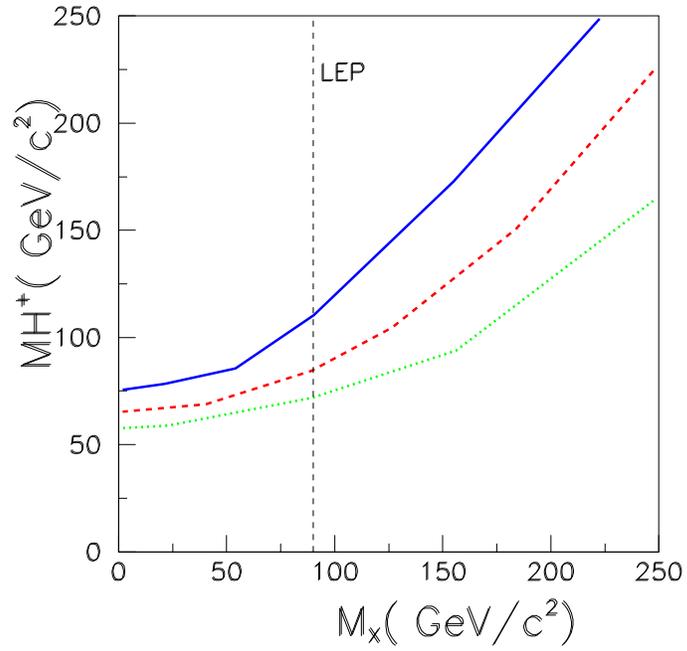,height=9cm,width=0.6\textwidth}}}
\caption{Lower limit of the charged Higgs boson mass as a function of
the lightest chargino mass for $B(b \to s\gamma)$ compatible with CLEO
measurement in the MSSM (solid) and in MSSM--BRpV (dashes and dots as
explained in the text).  The vertical dashed line corresponds to
$m_{\chi_1}=90$ GeV.}
\label{mhcha}
\end{figure}

\begin{figure}
\centerline{\protect\hbox{\psfig{file=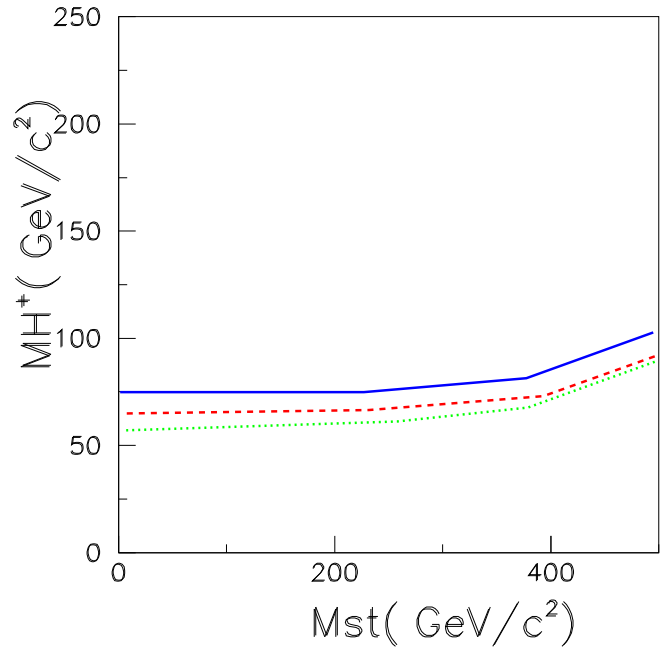,height=9cm,width=0.6\textwidth}}}
\caption{Lower limit of the charged Higgs boson mass as a function of
the lightest stop mass for $B(b \to s\gamma)$ compatible with CLEO
measurement in the MSSM (solid) and in the MSSM--BRpV (dashes and dots
as explained in the text).}
\label{mhstp}
\end{figure}

In Fig.~\ref{mhcha} we give the lower bounds on $m_{H^{\pm}}$
as a function of the lightest chargino mass $m_{\chi^{\pm}_1}$. The 
solid curve corresponds to the MSSM limit extracted from 
Fig.~\ref{mhcha1} and the dotted curve corresponds to the MSSM--BRpV
limit extracted from Fig.~\ref{mhcha2}. If we define the charged Higgs 
boson as the massive field $S_i^{\pm}$ with largest component along the 
rotated charged Higgs fields $H'^{\pm}_1$ and $H'^{\pm}_2$, which 
decouple from the rotated staus fields as $v'_3 \to  0$, then we
find the limit represented by the dashed curve. We see from this last
curve that in order to have $m_{\chi^{\pm}_1}>90$ GeV 
compatible with $B(b \to  s\gamma)$ we need $m_{H^{\pm}}\gsim 85$ 
GeV, therefore, relaxing the MSSM bound by about 25 GeV. In the same way,
in Fig.~\ref{mhstp} we plot the same lower bounds on $m_{H^{\pm}}$ but 
this time as a function of the lightest stop mass $m_{\tilde t_1}$. We 
observe from this figure that in order to cancel large contributions
to $B(b \to  s\gamma)$ due to a light charged Higgs boson, it is
more important to have a light chargino rather than a light stop.

Now a word about the theoretical uncertainties on the calculation of
$B(b \to  s\gamma)$. If we assume a $10\%$ error, then the bound
on the charged Higgs boson mass in the heavy stop limit within the MSSM
reduces to $m_{H^{\pm}}\gsim 320$ GeV. For the same reason, the 
corresponding bounds on the MSSM--BRpV reduce to 
$m_{H^{\pm}}\gsim 200-250$ GeV, which corresponds to a decrease in
70--120 GeV, \ie, comparable to the values quoted above. No changes are
observed in the case of light charged Higgs limits.

In summary, we have proved that the bounds on the charged Higgs mass
of the MSSM coming from the experimental measurement of the branching
ratio $B(b \to s\gamma)$ are relaxed if we add a single bilinear
R--Parity violating term of the form $\epsilon_3\widehat L_3\widehat
H_2$ to the superpotential. This term induces a tau neutrino mass
which in models with universality of soft breaking mass parameters at
the unification scale is naturally small. We study the effect of BRpV
on $B(b \to s\gamma)$ by considering the unconstrained model where the
values of all the unknown parameters are free at the weak scale. In
this case the main constraint comes from the smallness of the tau
neutrino mass. Even though in the MSSM--BRpV model the tau lepton
mixes with charginos, implying that the tau-lepton also contributes to
$B(b \to s\gamma)$ in loops with up--type squarks, we have shown that
this contribution is negligible.

In contrast, in the MSSM--BRpV model the staus mix with the charged
Higgs bosons and these contribute importantly to $B(b \to s\gamma)$ in
loops with up--type quarks.  For squark masses of a few TeV, where the
chargino contribution is negligible, the charged Higgs mass in the
MSSM has to satisfy $m_{H^{\pm}}\gsim 440$ GeV. This bound in the
MSSM--BRpV turns out to be $m_{H^{\pm}}\gsim 340-370$ GeV, therefore,
relaxing it in about 70--100 GeV. In order to have a light charged
Higgs boson in SUSY, its large contribution to $B(b \to s\gamma)$ can
only be compensated by a large contribution from a light chargino and
squark. In order to satisfy the experimental bound on $B(b \to
s\gamma)$ with $m_{\chi^{\pm}_1}>90$ GeV in the MSSM it is necessary
to have $m_{H^{\pm}}\gsim 110$ GeV. In the MSSM--BRpV model this bound
is $m_{H^{\pm}}\gsim 75-85$ GeV, i.e. 25--35 GeV weaker than in the
MSSM.  It is important to note that, in contrast to the MSSM, charged
Higgs boson masses as small as these can be achieved in MSSM--BRpV
already at tree level, as discussed in ref. \cite{v3cha}.  The reason
to the relaxation of the MSSM bounds can be understood as follows:
while the charged Higgs couplings to quarks diminish with the presence
of Higgs--Stau mixing, the contribution from the staus not always
compensate this decrease because the stau mass is, in general,
different from the charged Higgs boson mass, and could be larger.

Finally, a last word on our results on Fig. 5, 8 and 9 represented by
the dashed and dotted curves. These denote the Higgs mass bounds we
have obtained in the MSSM-BRpV model, when different basis are chosen
to perform the calculation. The point to stress is that our results do
{\sl not} depend on the choice of basis as such. They depend only on
our {\sl criterium} for specifying which state corresponds to the
Higgs boson and it is here where we have suggested two possible
definitions which are motivated by two possible basis choices.

\section*{Acknowledgements}

We thank the Warsaw HEP group in general, and J. Rosiek in particular,
for sharing the fortran code for the QCD corrections to the $b \to
s\gamma$ branching ratio described in refs.~\cite{CMisiakM} and
~\cite{MPR}.  This work was supported by DGICYT grant PB95-1077 and by
the EEC under the TMR contract ERBFMRX-CT96-0090.  M.A.D. was
supported by a postdoctoral grant from Ministerio de Educaci\'on y
Ciencias.


\newpage



%

\end{document}